\documentclass[sort&compress]{elsarticle}
\usepackage{amsmath,mathtools,amssymb}
\usepackage{graphicx}   %插入图片
\usepackage{epstopdf}
\usepackage{subfigure}
\usepackage{color}
\usepackage{verbatim} 
\usepackage{booktabs}
\usepackage{bm}   %公式加粗
\usepackage{dutchcal}
\usepackage[left=3cm,right=3cm,top=2.5cm,bottom=2.5cm]{geometry}  %设置页边距
\usepackage{lineno,hyperref}
\usepackage{tablists} 
\usepackage{makecell} 
\usepackage{tikz}
\usepackage{hyperref}
\usepackage{pstricks}
\usepackage[titletoc]{appendix}
\usepackage{pstricks-add}
\usepackage{float}
\usepackage{mathrsfs} 
\usepackage{amsthm}
\hypersetup{
	colorlinks=ture,
	linkcolor=blue,
	filecolor=gray,
	urlcolor=blue,
	citecolor=blue}
\numberwithin{equation}{section}
\allowdisplaybreaks[4]

\begin{document}
	\begin{frontmatter}
		\title{Higher  Chern-Simons-Antoniadis-Savvidy forms based on crossed modules} 
	
		\author{Danhua Song\corref{cor1}}
		\ead{danhua_song@163.com}
		\author{Ke Wu}
		\ead{wuke@cnu.edu.cn}
		\author{Jie Yang}
		\ead{5972@cnu.edu.cn}
		\cortext[cor1]{Corresponding author.}
		\address{School of Mathematical Sciences, Capital Normal University, Beijing 100048, China}
		\date{}
		
		\begin{abstract} 
			We present higher Chern-Simons-Antoniadis-Savvidy (ChSAS) forms based on crossed modules. 
			We start from introducing a generalized multilinear symmetric invariant polynomial for the differential crossed modules and
			constructing a metric independent, higher gauge invariant, and closed form using the higher curvature forms. 
		    Then, we establish the higher Chern-Weil theorem and prove that the higher ChSAS forms are a special case of this theorem.
		    Finally, we get the link of  two independent higher ChSAS theories.
			
		\end{abstract}
		
	\end{frontmatter}
	%
	
	%	\tableofcontents

	\section{Introduction}
  %higher order polynomials of the curvature forms.
	
	Higher gauge theory \cite{Baez.2010, FH, Baez2005HigherGT, 	JBUS, Faria_Martins_2011, doi:10.1063/1.4870640}  is a generalization of ordinary gauge theory, 
	where the underlying algebraic structure is promoted from an ordinary group to a higher group. 
	%It is believed that the higher gauge theory describes the dynamics of higher dimensional extended objects.
	And the associated gauge potentials and their gauge curvatures are generalized to higher degree forms valued in higher algebras.
	In this paper, we need only consider the case of 2-gauge theory. In conventional gauge theory, the connection 1-form equips curves with holonomies  in the gauge group $G$. While, in the 2-gauge theory, there is a 2-connection consisting of the differential 1-form and 2-form, which can be used to equip
surfaces with a new kind of surface holonomy, given by elements of another group $H$. 
More precisely, the algebraic
structure that replaces the gauge group in  2-gauge theory is a crossed module
 $(H, G; \alpha, \vartriangleright)$, as described in section \ref{sub1} below.
	
	Recently, the 2-gauge theory has been deeply studied  and applied with great interests. 
	For instance, the higher electromagnetic  \cite{HP, MHCT}, Yang-Mills \cite{2002hep.th....6130B, SDH111} and BF theories \cite{Martins:2010ry, FGHEMP}  have been developed under the framework of the 2-gauge theory.
	 In Ref. \cite{ SDH4}, the authors constructed 2-Chern-Simons (2CS) and 3-Chern-Simons (3CS) gauge theories by applying the Generalized Differential Calculus (GDC) \cite{Robinson1}.  Soncini and Zucchini  formulated a 4-dimensional  semistrict higher gauge theoretic Chern-Simons (CS) theory in Refs. \cite{ESRZ, R.Z, R.Z2, R.Z3, R.Z4, R.Z5}. Moreover, the higher Chern-Simons theories arising in the AKSZ-formalism were studied based on the $L_\infty$-algebra in literature  \cite{HSUSJS, DFUSJS, DFCLRUS, PRCS}.   
	The above mentioned works focus on the higher Chern-Simons theories in low dimension. While in this paper, the  higher ChSAS  (2n+2)-form based on the crossed modules is constructed, which can reduce to the 4-dimensional 2CS form in Ref. \cite{ SDH4}.

	In the ordinary Chern-Simons gauge theory, 
	the Chern-Simons action is constructed by
	the connection 1-form $A$ on a principle $G$-bundle, and the equation of motion implies the flatness of the connection, i.e., the corresponding curvature 2-form
$F=dA + A \wedge A$ vanishes.
The associated $(2n+2)$-Pontryagin-Chern form $P_{2n+2}$ is a higher order polynomial of the curvature form which is given by
\begin{align}\label{PC}
	P_{2n+2}=\langle F^{n+1}\rangle_{\mathcal{g}},
\end{align}
where $\mathcal{g}$ is the Lie algebra of $G$ and $\langle \cdots \rangle_{\mathcal{g}}$ stands for a multilinear symmetric invariant polynomial 
$
\langle \cdots \rangle_{\mathcal{g}}: \mathcal{g}^{n+1}\longrightarrow \mathbb{R}$  (see Ref. \cite{FIP}).
It is straightforward to show that $P_{2n+2}$ is a closed form, i.e., $d P_{2n+2} =0$. By the Poincar\'e lemma, there exists a $(2n+1)$-form $C^{2n+1}$ locally such that $P_{2n+2}=d C^{2n+1}$, which is called a Chern-Simons form. 
%	satisfying 
%	\begin{align}\label{sy}
%		\langle X_1 \cdots [X, X_i] \cdots X_j \cdots X_{n+1}\rangle_{\mathcal{g}}= -\langle X_1\cdots X_i \cdots [X, X_j]\cdots X_{n+1}\rangle_{\mathcal{g}}.
%	\end{align}
%	The symmetry of the polynomial essentially boils down to
%	\begin{align}
%		\langle D(A_1 \wedge \cdots \wedge A_{n+1})\rangle_{\mathcal{g}}=d\langle A_1 \wedge \cdots \wedge A_{n+1}\rangle_{\mathcal{g}},
%	\end{align}
%	where $\{ A_i, i=1, \cdots, n+1\}$ is a set of $\mathcal{g}$-valued differential forms.
	Besides, the explicit expression for the Chern-Simons form can be obtained as a special case of the Chern-Weil theorem.
	
	\textbf{Chern-Weil theorem}: 
	For two $\mathcal{g}$-valued connection $1$-forms $A_1$ and $A_0$ and the corresponding curvatures $F_1$ and $F_0$, there is an interpolation 
$A_t= A_0 + t (A_1 - A_0), (0 \leq t \leq 1)$,
	and the corresponding curvature is 
	$F_t = dA_t + A_t \wedge A_t$.
	Then, we  have
	\begin{align}\label{PCF}
		\langle F^{n+1}_1 \rangle_{\mathcal{g}} - 	\langle F^{n+1}_0 \rangle_{\mathcal{g}}= d Q^{2n+1}(A_0, A_1),
	\end{align}
	where 
	\begin{align}\label{transgression}
		Q^{2n+1}(A_0, A_1)=(n+1)\int_{0}^{1}dt \langle (A_1 - A_0) \wedge F^n_t\rangle_{\mathcal{g}}
	\end{align}
	is called a transgression $(2n+1)$-form.
	
	Setting $A_0=0$ and $A_1=A$ in \eqref{transgression}, we get 
	\begin{align}
		C^{2n+1}=Q^{2n+1}(0, A)=(n+1)\int_{0}^{1}dt \langle A \wedge (t dA + t^2 A \wedge A)^n\rangle_{\mathcal{g}}.
	\end{align}
%It is straightforward to show that the Chern–Simons forms are quasi-invariant under gauge transformations.

	Recently,
	 Antoniadis,  Konitopoulos and Savvidy   introduced a closed  invariant form similar to the Pontryagin-Chern form  \eqref{PC}  in the context of the so-called
	extended gauge theory, denoted by $\Gamma_{2n+p}$ with $p=3, 4, 6, 8$ (see Refs. \cite{GS, IAGS, GS1, SKGS}). Since $d \Gamma_{2n+p}=0$, one can have $ \Gamma_{2n+p}= d \mathfrak{C}^{(2n+p-1)}_{ChSAS}$, where $\mathfrak{C}^{(2n+p-1)}_{ChSAS}$ is just a ChSAS $(2n+p-1)$-form. And the ChSAS form can be expressed explicitly in terms of a higher order polynomial of the curvature forms. As with the standard Chern–Simons form, the ChSAS form is  background-free,
quasi-invariant and only locally defined.
	Later, P. Salgado and S. Salgado found that the extended  invariant form $\Gamma_{2n+p}$ can  also be constructed from an algebraic structure known as free differential algebra in Refs. \cite{SS, PSSS}.
	
	Especially, in Refs. \cite{FIP, FIPSPSS}, Salgado et al. studied a particular extended  invariant form which is given by
	\begin{align}\label{ei}
		\Gamma_{2n+3}= \langle F^n \wedge H\rangle_{\mathcal{g}},
	\end{align}
	where $H=d B + [A,B]$ is the $3$-form field-strength for the $2$-form field $B$ valued in $\mathcal{g}$. It is simple to prove that $d \Gamma_{2n+3}=0$, seeing the proof in Ref. \cite{IAGS}.
	Then, there is a $(2n+2)$-ChSAS form $\mathfrak{C}^{2n+2}_{ChSAS}$ satisfying $ \Gamma_{2n+3}= d\mathfrak{C}^{2n+2}_{ChSAS}$.
	Besides, they found that the explicit expression for $\mathfrak{C}^{2n+2}_{ChSAS}$  can be obtained as a special case of the generalized Chern-Weil theorem.
	
	\textbf{Generalized Chern-Weil theorem}: For two generalized connection forms $(A_0, B_0)$ and $(A_1, B_1)$ and their curvatures $(F_0, H_0)$ and $(F_1, H_1)$, there are two interpolations $A_t= A_0 + t (A_1- A_0)$ and $B_t= B_0 + t (B_1- B_0)$,  $(0 \leq t \leq 1)$, and the corresponding curvatures $F_t = d A_t + A_t \wedge A_t$ and $H_t= dB_t + [A_t, B_t]$. Then, have 
	\begin{align}\label{GCW}
		\langle F_1^n \wedge H_1\rangle_{\mathcal{g}}-
			\langle F_0^n \wedge H_0\rangle_{\mathcal{g}}=d Q^{2n+2}(A_0, B_0; A_1, B_1),
	\end{align}
	where
	\begin{align}\label{AST}
		Q^{2n+2}(A_0, B_0; A_1, B_1)= \int_{0}^{1}dt\Big\{n\langle F^{n-1}_t \wedge (A_1- A_0) \wedge H_t \rangle_{\mathcal{g}} + \langle F^n_t \wedge (B_1- B_0) \rangle_{\mathcal{g}}\Big\}
	\end{align}
	is called an ``Antoniadis-Savvidy Transgression " (AST) form.
	
	Setting $A_0=B_0=0$, $A_1=A$ and $B_1=B$ in \eqref{AST}, we can get the $(2n+2)$-ChSAS form
	\begin{align}\label{CASA}
		\mathfrak{C}^{2n+2}_{ChSAS}= Q^{2n+2}(0,0; A, B)=\int_{0}^{1}dt \langle n A \wedge F_t^{n-1}\wedge H_t + F^n_t \wedge B \rangle_{\mathcal{g}}.
	\end{align}

From a higher gauge theory of view, it is natural to ask a question: whether the above results are adapted to the case of  2-connections?
	It is the purpose of this paper to show that the extended  invariant form \eqref{ei} is well adapted to generalize the higher Chern-Simons considered in Ref. \cite{SDH4}.
	
This paper is organized as follows. In section \ref{sub1}, we briefly review the relevant topics of the 2-gauge theories and introduce the Lie algebra-valued differential forms and some conventions presented in this paper.
In section \ref{sub2}, we define a generalized multilinear symmetric invariant polynomial for the differential crossed modules. Then, we construct  a higher invariant form consisting of the higher curvatures and find the associated higher ChSAS form.
In section \ref{sub3}, we extend the generalized Chern-Weil theorem to the higher Chern-Weil theorem and prove that the results in the section \ref{sub2} can be obtained as a special case of the higher Chern-Weil theorem.
In section \ref{sub4}, we consider a kind of higher transgression gauge field theory where the lagrangian is a higher AST form. Finally, we know that two independent higher ChSAS theories have closed links under some boundary conditions.

	\section{2-gauge theory}\label{sub1}
	Let us begin by giving a short review of the $2$-gauge theory. For additional information on the topic, see for example Refs. \cite{Baez.2010, FH, Baez2005HigherGT, 	JBUS}.
	
	Firstly, the relevant algebraic tool involved in the description of this higher theory is a Lie crossed module $(H, G; \bar{\alpha}, \bar{\vartriangleright})$, which consists of two Lie groups $H$ and $G$ together with a smooth morphism $\bar{\alpha}: H \longrightarrow G$ and a smooth left action $ \bar{\vartriangleright}$ of $G$ on $H$ by automorphisms such that 
	\begin{align}
		\bar{\alpha}(g\bar{\vartriangleright} h)= g\bar{\alpha}(h)g^{-1},\ \ \ 
		\bar{\alpha}(h)\bar{\vartriangleright}  h'=h h' h^{-1},
	\end{align}
	for each $g \in G$ and $h, h' \in H$. For the crossed module $(H, G; \bar{\alpha}, \bar{\vartriangleright})$, the corresponding infinitesimal version is the differential crossed module $(\mathcal{h}, \mathcal{g}; \alpha, \vartriangleright)$, which  consists of two Lie algebras $\mathcal{g}:=$Lie $G$ and $\mathcal{h}:=$Lie $H$ together with a Lie algebra map $\alpha: \mathcal{h} \longrightarrow \mathcal{g}$ and a left action $\vartriangleright$ of $\mathcal{g}$ on $\mathcal{h}$ by derivations such that
	\begin{align}
		\alpha(X \vartriangleright Y)=[X, \alpha(Y)],\ 
		\alpha(Y)\vartriangleright Y'=[Y, Y'],
	\end{align}
	for each $X \in \mathcal{g}$ and $Y, Y' \in \mathcal{h}$.
%	In order not to
%introduce additional notations, we use the same letters $\alpha$ and $\vartriangleright$ for counterparts in the
%infinitesimal version of the crossed module.
	
	Thanks to the fact that the higher gauge fields are viewed as higher differential forms valued in the higher algebras, we consider the algebra-valued differential forms and introduce the component notations.
	Let $\Lambda^k(M, \mathcal{g})$ be the vector space of $\mathcal{g}$-valued differential $k$-forms on the manifold $M$ over $C^{\infty}(M)$. For $A \in \Lambda^k(M, \mathcal{g})$, have $A= \sum\limits_{a}A^a X_a$ with a scalar differential $k$-form $A^a$ and a basis $X_a$ for $\mathcal{g}$. Here, we consider matrix Lie algebras, and have $[X, X']= X X' -X'X$ for each $X, X' \in \mathcal{g}$. Define
\begin{align}
 dA := \sum\limits_{a} d A^a X_a,\ 
A_1 \wedge A_2 := \sum \limits_{ab}A^a_1 \wedge A^b_2 X_a X_b,\\
A_1 \wedge^{[, ]}A_2 := \sum \limits_{ab}A^a_1 \wedge A^b_2 [X_a, X_b],
\end{align}
for $A_1 = \sum\limits_{a} A^a_1X_a \in \Lambda^{k_1}(M, \mathcal{g})$, $A_2 = \sum\limits_{a} A^a_2X_a \in \Lambda^{k_2}(M, \mathcal{g})$.
Then there is an identity
\begin{align}
	A_1 \wedge^{[, ]}A_2 = A_1 \wedge A_2 - (-1)^{k_1 k_2}A_2 \wedge A_1.
\end{align}
The convention also applies to $\mathcal{h}$. Besides, for $B=\sum \limits_{b}  B^b Y_b \in \Lambda^{t}(M, \mathcal{h})$ with $Y_b \in \mathcal{h}$, one can define
\begin{align}
	\alpha(B):= \sum \limits_{b}B^b \alpha(Y_b),\ 
	A \wedge^{\vartriangleright}B:= \sum \limits_{ab} A^a \wedge B^b X_a \vartriangleright Y_b.
\end{align}

Besides, given a $\mathcal{g}$-valued connection 1-form $A$, one can define a `covariant derivative' $D$ on a $\mathcal{g}$-valued differential form $A'$ and an $\mathcal{h}$-valued differential form $B$, respectively,
\begin{align}
  D A'= d A' + A \wedge^{[, ]}A', \ DB= dB + A \wedge^{\vartriangleright}B.
\end{align}

Then, we consider the basic gauge fields of 2-gauge theory, which are given by  2-connections. Given a crossed module $(H, G; \bar{\alpha}, \bar{\vartriangleright})$ with the associated differential crossed module $(\mathcal{h}, \mathcal{g}; \alpha, \vartriangleright)$, the 2-connection $(A, B)$ is described by a $\mathcal{g}$-valued 1-form $A$  and an $\mathcal{h}$-valued 2-form $B$.
The corresponding $\mathcal{g}$-valued fake curvature 2-form and $\mathcal{h}$-valued 2-curvature 3-form are given by
\begin{align}\label{cur}
	\mathcal{F}= dA + \dfrac{1}{2}A \wedge^{[, ]} A -\alpha(B),\ 
\mathcal{G}= dB + A \wedge^{\vartriangleright}B,
\end{align}
which automatically satisfy the 2-Bianchi Identities:
\begin{align}
	&d \mathcal{F} + A \wedge^{[, ]}\mathcal{F} + \alpha(\mathcal{G})=0,\label{2bi}\\
	& d \mathcal{G} + A \wedge^{\vartriangleright}\mathcal{G} - \mathcal{F}\wedge^{\vartriangleright}B=0.\label{2bi1}
\end{align}
 We call $(A, B)$ fake-flat, if $\mathcal{F}=0$, and flat, if it is fake-flat and $\mathcal{G}=0$.
 
Moreover, there is a general 2-gauge transformation for the 2-connection $(A, B)$:
\begin{align}
&A'=g^{-1}Ag + g^{-1}dg + \alpha(\phi),\label{2gt1}\\
&B'=g^{-1}\vartriangleright B + d \phi + A'\wedge^{\vartriangleright}\phi - \phi\wedge\phi,\label{2gt2}
\end{align}
with $g \in G$ and $\phi \in \Lambda^1(M, \mathcal{h})$. The corresponding curvatures transform as follows:
\begin{align}
	\mathcal{F}'= g^{-1}\mathcal{F}g,\ 
   \mathcal{G}'= g^{-1}\vartriangleright \mathcal{G} + \mathcal{F}' \wedge^{\vartriangleright}\phi.
\end{align}

\section{Higher invariant and ChSAS forms}\label{sub2}
Based on the 2-gauge theory, we now turn to the construction of the higher invariant forms and find the explicit expression of the associated ChSAS forms.
A key issue for the construction is the establishment of the higher order polynomial of the higher curvature forms.

Inspired by the invariant polynomial $\langle \cdots \rangle_{\mathcal{g}}$ for the Lie algebra $\mathcal{g}$, we can define a generalized multilinear symmetric invariant polynomial for the differential crossed modules $(\mathcal{h}, \mathcal{g}; \alpha, \vartriangleright)$:
\begin{align}\label{mp}
	\langle \cdots, \cdot \rangle_{\mathcal{g}\mathcal{h}}: \mathcal{g}^n \times \mathcal{h}\longrightarrow \mathbb{R},
\end{align}
satisfying
\begin{align}
	&\langle X_1 \cdots  X_i \cdots X_n, X \vartriangleright Y\rangle_{\mathcal{g}\mathcal{h}}=-\sum_{i=1}^{n} \langle X_1 \cdots  [X, X_i] \cdots X_n, Y\rangle_{\mathcal{g}\mathcal{h}},\label{sy2}\\
	&\langle X_1 \cdots \alpha(Y_i) \cdots X_n, Y \rangle_{\mathcal{g}\mathcal{h}} = \langle X_1 \cdots \alpha(Y) \cdots X_n, Y_i\rangle_{\mathcal{g}\mathcal{h}}.\label{sy3}
\end{align}
%which are a generalization of the $G$-invariant form $\langle \cdot, \cdot \rangle_{\mathcal{g}\mathcal{h}}$ on $(\mathcal{g}, \mathcal{h}; \alpha, \vartriangleright)$ in Ref.  \cite{SDH4}.

The symmetry implies that 
\begin{align}\label{sy4}
	\langle X_1\cdots X_i \cdots X_j \cdots X_n, Y \rangle_{\mathcal{g}\mathcal{h}}=\langle X_1\cdots X_j \cdots X_i \cdots X_n, Y \rangle_{\mathcal{g}\mathcal{h}},
\end{align}
and the invariance states clearly that
\begin{align}\label{inv}
	\langle g X_1g^{-1} \cdots g X_n g^{-1},  g\vartriangleright Y\rangle_{\mathcal{g}\mathcal{h}}=\langle X_1\cdots X_n, Y \rangle_{\mathcal{g}\mathcal{h}},
\end{align}
for each $g \in G$, which can be given by taking $g$ as an infinitesimal transformation and using the identity \eqref{sy2}.
	In the case of $n=1$, \eqref{mp} becomes a bilinear form $\langle \cdot, \cdot \rangle_{\mathcal{g}\mathcal{h}} : \mathcal{g}\times \mathcal{h}\longrightarrow \mathbb{R}$ in \cite{SDH4, R.Z5}.

 For example, there is a crossed module $(T, T; \bar{\alpha}, \bar{\vartriangleright})$, where $T$ is a compact connected abelian group, $\bar{\alpha}: T \longrightarrow T$ is an endomorphism and $\bar{\vartriangleright}: T \times T \longrightarrow T$ is the trivial action of $T$ on itself. The associated differential crossed module is $(\mathcal{t}, \mathcal{t}; \alpha, \vartriangleright)$, where the algebra $\mathcal{t}$ is abelian and the action $\vartriangleright$ is trivial. Let $\langle \cdot, \cdot \rangle_{\mathcal{t}}$ be a symmetric non singular bilinear form on $\mathcal{t}$  and the endomorphism $\alpha$ be symmetric with respect to this pairing. See \cite{R.Z5} for details.

 	Moreover, there is a Poincar\'e 2-group $(H=\mathbb{R}^3, G=SO(2,1); \bar{\alpha}, \bar{\vartriangleright})$, where $\bar{\alpha}$ is trivial and $\bar{\vartriangleright}$ is the representation of $SO(2,1)$ on $\mathbb{R}^3$. The corresponding Poincar\'e 2-algebra is $(\mathcal{h}=\mathbb{R}^3, \mathcal{g}=so(2,1); \alpha, \vartriangleright)$ with $\alpha(P^a)=0$ for $P^a \in \mathcal{h}$ and $J^i \vartriangleright P^a=\epsilon^{ia}_{\ \ b} P^{b}$, where 
 $\{ J^i, i=0,1,2 \}$ and $\{ P^a, a=0,1,2 \}$ are the generators of $so(2,1)$ and $\mathbb{R}^3$, respectively. 
 The commutation relation is $[ J^i, J^j]= \epsilon^{ijk}J_k$, and $\epsilon^{ijk}$ is the Levi-Civita symbol.
 Then,  \eqref{mp} can be given by 
 \begin{align}\label{a}
 	\langle J^i, P^a \rangle_{\mathcal{g}\mathcal{h}}= \eta^{i,a}
 \end{align}
    for the case of $n=1$, where $\eta = diag(-, +, +)$. 
    It is easy to prove that the pairing satisfies \eqref{sy2} and \eqref{sy3}.
    The above algebra with the pairing \eqref{a} has been used for generalizing (2+1)-dimensional pure gravity with
vanishing cosmological constant to the higher gauge theory formalism in \cite{RBEM}.

	More examples can be found in the works of  Baez, Schreiber \cite{JADU} and Zucchini \cite{R.Z5}. 
	Generally, for the differential crossed modules $(\mathcal{h}, \mathcal{g}; \alpha, \vartriangleright)$, where $\mathcal{g}$ is a matrix Lie algebra, the generalized multilinear symmetric invariant polynomial can be given by
	\begin{align}
	\langle X_1 \cdots X_n, Y_{n+1} \rangle_{\mathcal{g}, \mathcal{h}}=& \sum_{i_1,\cdots, i_{n}} \Big\{ tr(\alpha(Y_{n+1})X_{i_1} \cdots X_{i_n})+ tr(X_{i_1}\alpha(Y_{n+1})X_{i_2} \cdots X_{i_n})\nonumber\\
	&+ \cdots + tr(X_{i_1} \cdots X_{i_n}\alpha(Y_{n+1}))\Big\}.
\end{align}

Besides, the equation \eqref{sy2} essentially boils down to
\begin{align}\label{Dd}
	\langle D(A_1\wedge \cdots \wedge A_n, \hat{B})\rangle_{\mathcal{g}\mathcal{h}}=d \langle A_1\wedge \cdots \wedge A_n, \hat{B}\rangle_{\mathcal{g}\mathcal{h}},
\end{align}
where $\{ A_i, i=1, \cdots, n\}$ is a set of $\mathcal{g}$-valued differential forms and $\hat{B}$ is an $\mathcal{h}$-valued differential form.

Thus, imitating the extended invariant form \eqref{ei}, we can construct a higher invariant form based on the 2-gauge field theory
\begin{align}
	\mathcal{P}_{2n+3}=\langle \mathcal{F}^n, \mathcal{G}\rangle_{\mathcal{g}\mathcal{h}},
\end{align}
where $\mathcal{F}$ and $\mathcal{G}$ are the higher curvature forms \eqref{cur}.
It is possible to prove that the higher pairing form is also 2-gauge invariant
under the 2-gauge transformation \eqref{2gt1} and \eqref{2gt2} by direct computations.
Additionally, using eqs. \eqref{sy3}, \eqref{Dd} and the 2-Bianchi-Identities \eqref{2bi}-\eqref{2bi1}, we have
\begin{align}
 d \langle \mathcal{F}^n, \mathcal{G}\rangle_{\mathcal{g}\mathcal{h}}&=
 \langle D(\mathcal{F}^n, \mathcal{G} )\rangle_{\mathcal{g}\mathcal{h}}\nonumber\\
 &=n\langle D\mathcal{F} \wedge \mathcal{F}^{n-1}, \mathcal{G}\rangle_{\mathcal{g}\mathcal{h}}  + \langle \mathcal{F}^n, D \mathcal{G}\rangle_{\mathcal{g}\mathcal{h}}\nonumber\\
 &= -n\langle \alpha(\mathcal{G})\wedge \mathcal{F}^{n-1}, \mathcal{G}\rangle_{\mathcal{g}\mathcal{h}} + \langle \mathcal{F}^n, \mathcal{F}\wedge^{\vartriangleright}B\rangle_{\mathcal{g}\mathcal{h}}=0,
\end{align}
which means that $\mathcal{P}_{2n+3}$ is a closed form. 
By the Poincar\'e lemma, $\mathcal{P}_{2n+3}$ can be locally written as an exterior differential of a certain $(2n+2)$-form.

Actually, this  $(2n+2)$-form potential is  a categorical generalization of the ChSAS form in \eqref{CASA}. 
We find the certain $(2n+2)$-form by using a kind of variational approach in \cite{FIP}. Since
\begin{align}
	\delta\mathcal{F}= D \delta A - \alpha(\delta B),\ \  \delta \mathcal{G}= D \delta B + \delta A \wedge^{\vartriangleright}B,
\end{align}
the variation $\delta \mathcal{P}_{2n+3}$ is given by
\begin{align}\label{p}
	\delta \mathcal{P}_{2n+3}&= n \langle (D\delta A - \alpha(\delta B))\wedge \mathcal{F}^{n-1}, \mathcal{G}\rangle_{\mathcal{g}\mathcal{h}} + \langle \mathcal{F}^{n}, D \delta B + \delta A \wedge^{\vartriangleright}B \rangle_{\mathcal{g}\mathcal{h}}\nonumber\\
	&=d (n \langle \delta A \wedge \mathcal{F}^{n-1}, \mathcal{G}\rangle_{\mathcal{g}\mathcal{h}} + \langle \mathcal{F}^n, \delta B\rangle_{\mathcal{g}\mathcal{h}}).
\end{align}
Following Ref. \cite{FIP}, we can introduce a one-parameter family of the higher potentials and strengths through the parameter $t$, $0\leq t \leq 1$:
\begin{align}
	&A_t=tA, \ \ \mathcal{F}_t= t\mathcal{F} + (t^2-t)A \wedge A;\\
	&B_t=tB, \ \ \mathcal{G}_t =t \mathcal{G} + (t^2-t)A \wedge^{\vartriangleright}B.
\end{align}

Choosing a variation of the form $\delta =\delta t (\partial/\partial t)$, we have $\delta A_t=\delta t A$ and $\delta B_t=\delta t B$. From \eqref{p}, we get
\begin{align}
	\mathcal{P}_{2n+3} = d \mathcal{C}^{2n+2}_{2ChSAS}
\end{align}
with
\begin{align}
	\mathcal{C}^{2n+2}_{2ChSAS}= \int_{0}^{1}dt \Big\{n\langle A \wedge \mathcal{F}^{n-1}_t, \mathcal{G}_t \rangle_{\mathcal{g}\mathcal{h}} + \langle \mathcal{F}^n_t, B \rangle_{\mathcal{g}\mathcal{h}}\Big\},
\end{align}
which we call a ``2-Chern-Simons-Antoniadis-Savvidy"   (2ChSAS) form.
From the above facts, we know that the extension in Ref. \cite{FIP} is applicable equally to the higher gauge field theory.

\section{Higher Chern-Weil theorem}\label{sub3}
In this subsection, we generalize the AST form \eqref{AST} and the generalized Chern-Weil theorem based on the 2-gauge theory. Let $(A_0, B_0)$ and $(A_1, B_1)$ be two 2-connections, and the corresponding curvature forms are given by
\begin{align}
 \mathcal{F}_i= dA_i+ \dfrac{1}{2}A_i \wedge^{[, ]} A_i - \alpha(B_i), \ \ \ 
	\mathcal{G}_i= dB_i + A_i \wedge^{\vartriangleright}B_i,
\end{align}
for $i=0,1$.
Define the interpolations between the two connections as
\begin{align}
	 A_t= A_0 + t \theta, \ \theta= A_1-A_0, \label{d1}\\
	 B_t = B_0 + t \Phi,\  \Phi= B_1 - B_0,\label{d2}
\end{align}
for $0 \leq t \leq 1$ and their curvatures are given by
\begin{align}\label{c1}
	\mathcal{F}_t = dA_t + \dfrac{1}{2}A_t \wedge^{[, ]} A_t - \alpha(B_t),\
	\mathcal{G}= dB_t + A_t \wedge^{\vartriangleright}B_t.
\end{align}
By direct computations, we have
\begin{align}\label{t}
	\dfrac{d\mathcal{F}_t}{dt}= D_t \theta - \alpha(\Phi),\ \ \dfrac{d\mathcal{G}_t}{dt}= D_t \Phi +\theta \wedge^{\vartriangleright}B_t,
\end{align}
where $D_t \theta = d \theta + A_t \wedge^{[, ]}\theta$ and $D_t \Phi = d \Phi + A_t \wedge^{\vartriangleright} \Phi$.

Then, the difference between $\mathcal{P}^{(1)}_{2n+3}$ and $\mathcal{P}^{(0)}_{2n+3}$ is an exact form,
\begin{align}\label{ch1}
\mathcal{P}^{(1)}_{2n+3}-\mathcal{P}^{(0)}_{2n+3}=\langle \mathcal{F}^n_1, \mathcal{G}_1 \rangle_{\mathcal{g}\mathcal{h}} - \langle \mathcal{F}^n_0, \mathcal{G}_0 \rangle_{\mathcal{g}\mathcal{h}}= d \mathcal{Q}^{2n+2}(A_0, B_0; A_1, B_1),
\end{align}
where
\begin{align}\label{2AST}
	\mathcal{Q}^{2n+2}(A_0, B_0; A_1, B_1)= \int_{0}^{1}dt\Big\{n\langle \theta \wedge \mathcal{F}^{n-1}_t, \mathcal{G}_t\rangle_{\mathcal{g}\mathcal{h}} + \langle \mathcal{F}^n_t, \Phi\rangle_{\mathcal{g}\mathcal{h}}\Big\}
\end{align}
is what we call a ``2-Antoniadis-Savvidy Transgression"(2AST) form.
\begin{proof}
	By \eqref{t}, we can compute
	\begin{align}\label{diff}
		&\langle \mathcal{F}^n_1, \mathcal{G}_1 \rangle_{\mathcal{g}\mathcal{h}} - \langle \mathcal{F}^n_0, \mathcal{G}_0 \rangle_{\mathcal{g}\mathcal{h}}
		= \int_{0}^{1}dt \frac{d}{dt}	\langle \mathcal{F}^n_t, \mathcal{G}_t \rangle_{\mathcal{g}\mathcal{h}} \nonumber\\
		&= \int_{0}^{1}dt \Big\{n\langle \mathcal{F}^{n-1}_t\wedge \frac{d\mathcal{F}_t}{dt}, \mathcal{G}_t \rangle_{\mathcal{g}\mathcal{h}} + \langle \mathcal{F}^n_t, \frac{d\mathcal{G}_t}{dt}\rangle_{\mathcal{g}\mathcal{h}}\Big\}\nonumber\\
		&= \int_{0}^{1}dt \Big\{n\langle \mathcal{F}^{n-1}_t\wedge (D_t \theta - \alpha(\Phi)), \mathcal{G}_t \rangle_{\mathcal{g}\mathcal{h}} + \langle \mathcal{F}^n_t, D_t \Phi +\theta \wedge^{\vartriangleright}B_t\rangle_{\mathcal{g}\mathcal{h}}\Big\}.
	\end{align}
Using  \eqref{sy2}, we have
\begin{align}\label{id}
	\langle \mathcal{F}^n_t, \theta \wedge^{\vartriangleright}B_t \rangle_{\mathcal{g}\mathcal{h}}=0.
\end{align}
%Then, we have
%\begin{align*}
%		\langle \mathcal{F}^n_1, \mathcal{G}_1 \rangle_{\mathcal{g}\mathcal{h}} - \langle \mathcal{F}^n_0, \mathcal{G}_0 \rangle_{\mathcal{g}\mathcal{h}}=  \int_{0}^{1}dt (n\langle \mathcal{F}^{n-1}_t\wedge (D_t \theta - \alpha(\Phi)), \mathcal{G}_t \rangle_{\mathcal{g}\mathcal{h}} + \langle \mathcal{F}^n_t, D_t \Phi \rangle_{\mathcal{g}\mathcal{h}}).
%\end{align*}
On the other hand, the RHS of \eqref{ch1} is given by
\begin{align}\label{diff1}
	&d\int_{0}^{1}dt\Big\{n\langle \theta \wedge \mathcal{F}^{n-1}_t, \mathcal{G}_t\rangle_{\mathcal{g}\mathcal{h}} + \langle \mathcal{F}^n_t, \Phi\rangle_{\mathcal{g}\mathcal{h}}\Big\}\nonumber\\
	&=\int_{0}^{1}dt \Big\{n \langle D_t(\theta \wedge \mathcal{F}^{n-1}_t,\mathcal{G}_t)\rangle_{\mathcal{g}\mathcal{h}} + \langle D_t( \mathcal{F}^n_t, \Phi)\rangle_{\mathcal{g}\mathcal{h}}\Big\}
\end{align}
Using \eqref{sy3} and the 2-Bianchi-Identities, we have
\begin{align}
	&\langle \theta \wedge D_t \mathcal{F}_t\wedge\mathcal{F}^{n-2}_t, \mathcal{G}_t\rangle_{\mathcal{g}\mathcal{h}}=- \langle \theta \wedge \alpha(\mathcal{G}_t)\wedge \mathcal{F}^{n-2}_t,\mathcal{G}_t \rangle_{\mathcal{g}\mathcal{h}}=0,\label{11}\\
	&\langle \theta \wedge\mathcal{F}^{n-1}_t, D_t\mathcal{G}_t \rangle_{\mathcal{g}\mathcal{h}}= \langle \theta \wedge \mathcal{F}^{n-1}_t, \mathcal{F}_t \wedge^{\vartriangleright}B_t\rangle_{\mathcal{g}\mathcal{h}}=0,\label{22}\\
	&\langle D_t \mathcal{F}_t \wedge \mathcal{F}^{n-1}, \Phi \rangle_{\mathcal{g}\mathcal{h}}=-\langle \mathcal{F}^{n-1}_t \wedge \alpha(\Phi), \mathcal{G}_t \rangle_{\mathcal{g}\mathcal{h}}.\label{33}
\end{align}
Substituting \eqref{id} into \eqref{diff} and \eqref{11}, \eqref{22}, \eqref{33} into \eqref{diff1}, we can conclude that
\begin{align*}
		\langle \mathcal{F}^n_1, \mathcal{G}_1 \rangle_{\mathcal{g}\mathcal{h}} - \langle \mathcal{F}^n_0, \mathcal{G}_0 \rangle_{\mathcal{g}\mathcal{h}}= d\int_{0}^{1}dt\Big\{n\langle \theta \wedge \mathcal{F}^{n-1}_t, \mathcal{G}_t\rangle_{\mathcal{g}\mathcal{h}} + \langle \mathcal{F}^n_t, \Phi\rangle_{\mathcal{g}\mathcal{h}}\Big\}.
\end{align*}
\end{proof}

The proof is similar in spirit to the proof of the ordinary Chern-Weil theorem. But, the related calculations are  completed based on the higher algebra structures.
From  \eqref{2AST}, we have for the case $A_0=B_0=0$ and $A_1=A$, $B_1=B$,
\begin{align}\label{Q}
	\mathcal{Q}^{2n+2}(0, 0; A, B)=\int_{0}^{1}dt \Big\{n \langle A \wedge \mathcal{F}^{n-1}_t, \mathcal{G}_t \rangle_{\mathcal{g}\mathcal{h}} + \langle \mathcal{F}^n_t, B \rangle_{\mathcal{g}\mathcal{h}}\Big\}= \mathcal{C}^{2n+2}_{2ChSAS}.
\end{align}
Particularly, for $n=1$, we set the same result in \cite{SDH4} where the author computed $Q_{2CS}$ as
 the $4$-d Chern-Simon form based on the crossed module 
\begin{align}
		\mathcal{Q}^{4}(0, 0; A, B)=\int_{0}^{1}dt \Big\{\langle A, \mathcal{G}_t \rangle_{\mathcal{g}\mathcal{h}} + \langle \mathcal{F}_t, B \rangle_{\mathcal{g}\mathcal{h}}\Big\}
		= Q_{2CS}.
\end{align}

\section{The higher Transgression Gauge Field Theory (TGFT)}\label{sub4}
In Ref. \cite{FIERPS}, there is a Transgression Gauge Field Theory (TGFT) whose lagrangian is an ordinary transgression form \eqref{transgression}. In stark contrast with the Chern-Simons gauge theory, the TGFT action is fully invariant rather than pseudo-invariant. In other wards, the TGFT theory has a gauge symmetry generated by a local gauge transformation.

A similar heuristic argument suggests that we can develop a higher TGFT theory,  where the 2AST form \eqref{2AST} is a lagrangian. And we can also consider the corresponding higher gauge symmetry generated by a local 2-gauge transformation \eqref{2gt1}-\eqref{2gt2}.

Consequently, the higher TGFT action can take the form
\begin{align}
	S^{2n+2}_{2ASTGFT}&= \int_{M} \mathcal{Q}^{2n+2}(A_0, B_0; A_1, B_1)\nonumber\\
	&=\int_{M} \int_{0}^{1}dt\Big\{n\langle \theta \wedge \mathcal{F}^{n-1}_t, \mathcal{G}_t\rangle_{\mathcal{g}\mathcal{h}} + \langle \mathcal{F}^n_t, \Phi\rangle_{\mathcal{g}\mathcal{h}}\Big\}.
\end{align}
Performing independent variations of $(A_1, B_1)$ and $(A_0, B_0)$ in the action, we have
\begin{align}\label{2ASTGFT}
	\delta S^{2n+2}_{2ASTGFT} = \int_{M} \Big\{n \langle \delta A_1 \wedge \mathcal{F}^{n-1}_1, \mathcal{G}_1\rangle_{\mathcal{g}\mathcal{h}} + \langle \mathcal{F}^n_1, \delta B_1\rangle_{\mathcal{g}\mathcal{h}} - n \langle \delta A_0 \wedge \mathcal{F}^{n-1}_0, \mathcal{G}_0\rangle_{\mathcal{g}\mathcal{h}} - \langle \mathcal{F}^n_0, \delta B_0\rangle_{\mathcal{g}\mathcal{h}} - d \hat{\Pi}\Big\},
\end{align}
with the boundary term
\begin{align*}
	\hat{\Pi}= n  \int_{0}^{1} dt \Big\{(n-1) \langle \theta \wedge \delta A_t \wedge \mathcal{F}^{n-2}_t, \mathcal{G}_t \rangle_{\mathcal{g}\mathcal{h}} + \langle \theta \wedge \mathcal{F}^{n-1}_t, \delta B_t \rangle_{\mathcal{g}\mathcal{h}} + \langle \delta A_t \wedge \mathcal{F}^{n-1}_t, \Phi \rangle_{\mathcal{g}\mathcal{h}}\Big\}.
\end{align*}

Read off the field equations from \eqref{2ASTGFT}
\begin{align}
	\langle \mathcal{F}^{n-1}_1, \mathcal{G}_1\rangle_{\mathcal{g}\mathcal{h}} =0, \ \ \ \langle \mathcal{F}^n_1, Y_b \rangle_{\mathcal{g}\mathcal{h}}=0;\label{eq1}\\
	\langle \mathcal{F}^{n-1}_0, \mathcal{G}_0\rangle_{\mathcal{g}\mathcal{h}} =0, \ \ \ \langle \mathcal{F}^n_0, Y_b \rangle_{\mathcal{g}\mathcal{h}}=0,\label{eq2}
\end{align}
where $Y_b$ is a basis for $\mathcal{h}$,
and the boundary condition is obtained by demanding the vanishing of $\hat{\Pi}$ on $\partial M$:
\begin{align}\label{bc}
    \Big\{ \int_{0}^{1} dt ((n-1) \langle \theta \wedge \delta A_t \wedge \mathcal{F}^{n-2}_t, \mathcal{G}_t \rangle_{\mathcal{g}\mathcal{h}} + \langle \theta \wedge \mathcal{F}^{n-1}_t, \delta B_t \rangle_{\mathcal{g}\mathcal{h}} + \langle \delta A_t \wedge \mathcal{F}^{n-1}_t, \Phi \rangle_{\mathcal{g}\mathcal{h}})\Big\}\Big|_{\partial M}=0.
\end{align}
Thus, from \eqref{eq1}-\eqref{bc}, we get that the two independent higher ChSAS theories with the 2ChSAS forms of $t=0$ and $t=1$ in \eqref{2AST}, are inextricably linked at the boundary.

Besides, under the 2-gauge transformations
\begin{align}
	&A'_1 = g^{-1}A_1 g + g^{-1}dg + \alpha(\phi),\label{2g1}\\
	&B'_1=g^{-1}\vartriangleright B_1 + d \phi + A'_1 \wedge^{\vartriangleright}\phi - \phi \wedge \phi,\label{2g2}\\
	&A'_0 = g^{-1}A_0 g + g^{-1}dg + \alpha(\phi),\label{2g3}\\
	&B'_0=g^{-1}\vartriangleright B_0 + d \phi + A'_0 \wedge^{\vartriangleright}\phi - \phi \wedge \phi,\label{2g4}
\end{align}
the corresponding differences of 2-connections in \eqref{d1}-\eqref{d2} and interpolation curvatures in \eqref{c1} change as
\begin{align*}
	\theta'= g^{-1}\theta g,\ \ \Phi'=g^{-1}\vartriangleright \Phi + \theta' \wedge^{\vartriangleright}\phi,\\
	\mathcal{F}'_t= g^{-1}\mathcal{F}_t g, \ \ \mathcal{G}'_t= g^{-1}\vartriangleright \mathcal{G}_t + \mathcal{F}'_t \wedge^{\vartriangleright}\phi.
\end{align*}
Using \eqref{inv}, we can prove that the higher TGFT action is fully 2-gauge invariant under \eqref{2g1}-\eqref{2g4}. Thus, we have actually proved that the higher TGFT theory has a 2-gauge symmetry.

\section{Concluding remarks}
We have shown in this work that: (i)
the extended invariant form \eqref{ei} is well adapted to the case of the  2-gauge theory;
(ii) the associated higher AST and ChSAS forms are also expressed explicitly and the procedure to find those expressions is analogous to the one described in Ref. \cite{FIP};
(iii) the generalized Chern-Weil theorem can be generalized to the higher Chern-Weil theorem;
 (iv) the higher TGFT action is fully 2-gauge invariant and two independent 2ChSAS theories are inextricably linked under some boundary conditions, analogous to the ordinary TGFT in Ref. \cite{FIERPS}.

In addition to the 2-gauge theory, the 3-gauge theory has been developed and applied based on the 2-crossed modules, seeing Refs. \cite{Faria_Martins_2011, doi:10.1063/1.4870640, Song, TRMV1, Radenkovic:2019qme}.
 Whether we can generalize these discussions in this paper to the 3-gauge theory, there may be a number of new things that should arise in this case.
Likewise, there may be higher gauge anomalies for these higher ChSAS forms under the higher gauge transformations. And there may be higher extension of Chern-Simons type characteristic classes and descent equations given in Refs. \cite{GWW1, GWW2}.
We leave these arguments for future work.

\section*{Acknowledgment}
This work is supported by the National Natural Science Foundation of China (Nos.11871350 and 11971322).

The authors would like to thank the anonymous referee and editor for their valuable comments and suggestions which helped us improve the paper.

\end{document}